\documentclass[epj]{webofc}
\usepackage[utf8]{inputenc}
\usepackage[varg]{txfonts}   
\usepackage{booktabs}
\usepackage{xcolor}
\definecolor{darkred}{rgb}{0.4,0.0,0.0}
\definecolor{darkgreen}{rgb}{0.0,0.4,0.0}
\definecolor{darkblue}{rgb}{0.0,0.0,0.4}
\usepackage[bookmarks,linktocpage,colorlinks,
    linkcolor = darkred,
    urlcolor  = darkblue,
    citecolor = darkgreen]{hyperref}
\def\slash#1{\mbox{$\not \!\! #1$}}
\def\Dslash{{\slash {\cal D}}}

\def\lvec#1{\setbox0=\hbox{$#1$}
    \setbox1=\hbox{$\scriptstyle\leftarrow$}
    #1\kern-\wd0\smash{
    \raise\ht0\hbox{$\raise1pt\hbox{$\scriptstyle\leftarrow$}$}}
    \kern-\wd1\kern\wd0}
\def\rvec#1{\setbox0=\hbox{$#1$}
    \setbox1=\hbox{$\scriptstyle\rightarrow$}
    #1\kern-\wd0\smash{
    \raise\ht0\hbox{$\raise1pt\hbox{$\scriptstyle\rightarrow$}$}}
    \kern-\wd1\kern\wd0}

\def\diracstar#1#2{
    \setbox0=\hbox{$\gamma$}\setbox1=\hbox{$\gamma_{#1}$}
    \gamma_{#1}\kern-\wd1\kern\wd0
    \smash{\raise4.5pt\hbox{$\scriptstyle#2$}}}

\def\tr{\,\hbox{tr}\,}

\newcommand{\beq}{\begin{equation}}
\newcommand{\eeq}{\end{equation}}
\newcommand{\beqn}{\begin{eqnarray}}
\newcommand{\eeqn}{\end{eqnarray}}

\graphicspath{{figures/}}

\newif\ifcomment
 \commentfalse

\usepackage{subfigure}
\wocname{EPJ Web of Conferences}
\woctitle{Lattice2017}
\begin{document}
%
\selectlanguage{english}
\title{Testing a non-perturbative mechanism for elementary fermion mass generation: numerical results
}
\author{%
\firstname{Stefano} \lastname{Capitani}\inst{1} \and
\firstname{Giulia Maria} \lastname{de Divitiis}\inst{2} \and
\firstname{Petros}  \lastname{Dimopoulos}\inst{2,3}\fnsep\thanks{Speaker, \email{dimopoulos@roma2.infn.it}} \and
\firstname{Roberto} \lastname{Frezzotti}\inst{2} \and
\firstname{Marco} \lastname{Garofalo}\inst{4} \and
\firstname{Bastian} \lastname{Knippschild}\inst{5} \and
\firstname{Bartosz} \lastname{Kostrzewa}\inst{5} \and
\firstname{Ferenc} \lastname{Pittler}\inst{5} \and
\firstname{Giancarlo} \lastname{Rossi}\inst{2,3} \and
\firstname{Carsten} \lastname{Urbach}\inst{5}
}
\institute{
Johann Wolfgang Goethe-Universit\"at Frankfurt am Main, Institut f\"ur Theoretische Physik, 
Max-von-Laue-Strasse 1 D-60438 Frankfurt am Main, Germany
\and
Dipartimento di Fisica, Università di Roma ``Tor Vergata" and INFN, Sezione di Roma 2, Via
della Ricerca Scientifica - 00133 Rome, Italy
\and
Centro Fermi - Museo Storico della Fisica e Centro Studi e Ricerche Enrico Fermi,
Compendio del Viminale, Piazza del Viminiale 1, I-00184, Rome, Italy
\and
Higgs Centre for Theoretical Physics, School of Physics and Astronomy, The University of
Edinburgh, Edinburgh EH9 3JZ, Scotland, UK
\and
Helmholtz Institut f\"ur Strahlen-und Kernphysik (Theorie), Nussallee 14-16 Bethe Center for Theoretical
Physics, Nussallee 12 Universität Bonn, D-53115 Bonn, Germany
}
\abstract{Based on a recent proposal according to which elementary particle masses could be generated by a 
non-perturbative dynamical phenomenon, alternative to the Higgs mechanism, we carry out lattice simulations of a model 
where a non-abelian strongly interacting fermion doublet is also coupled to a doublet of complex 
scalar fields via a Yukawa and an ``irrelevant" Wilson-like term. In this pioneering study we use naive 
fermions and work in the quenched approximation. 
We present preliminary numerical results both in the Wigner and in the Nambu-Goldstone phase, 
focusing on the observables relevant to check the occurrence of the 
conjectured dynamical fermion mass generation effect in the continuum limit of the critical theory in its spontaneously broken phase.
  
}
\maketitle
\section{Introduction}\label{intro}

In Refs.~\cite{Frezzotti:2014wja, Frezzotti:2013raa} a novel approach to the mass generation of elementary particles 
and the mass hierarchy problem has been proposed. It is based on a Non-Perturbative (NP) mechanism whose existence can be tested by studying,  
with the help of Lattice QCD (LQCD) simulations, the properties of a non-Abelian (SU(3) gauge) toy-model where 
an isospin doublet of strongly interacting fermions  is coupled to a complex scalar field via 
Yukawa and Wilson-like terms. The Lagrangian of the toy-model reads:

 \begin{eqnarray}
&&{\cal L}_{\rm{toy}}(\Psi,A,\Phi)= {\cal L}_{kin}(\Psi,A,\Phi)+{\cal V}(\Phi)
+{\cal L}_{Wil}(\Psi,A,\Phi) + {\cal L}_{Yuk}(\Psi,\Phi) \, ,\label{SULL} \\
&&\quad{\cal L}_{kin}(\Psi,A,\Phi)= \frac{1}{4}(F\cdot F)+
\bar \Psi_L\Dslash \Psi_L+\bar \Psi_R\Dslash \,\Psi_R+\frac{1}{2}{\tr}\big{[}\partial_\mu\Phi^\dagger\partial_\mu\Phi\big{]}\label{LKIN}\\
&&\quad{\cal V}(\Phi)= \frac{\mu_0^2}{2}{\tr}\big{[}\Phi^\dagger\Phi\big{]}+\frac{\lambda_0}{4}\big{(}{\tr}\big{[}\Phi^\dagger\Phi\big{]}\big{)}^2\label{LPHI}\\
&&\quad{\cal L}_{Yuk}(\Psi,\Phi)=\
  \eta\,\big{(} \bar \Psi_L\Phi \Psi_R+\bar \Psi_R \Phi^\dagger \Psi_L\big{)} \, , 
\label{Yukawa} \\
&&\quad{\cal L}_{Wil}(\Psi,A,\Phi)= \frac{b^2}{2}\rho\,\big{(}\bar \Psi_L
{\overleftarrow{\cal D}}_\mu\Phi {\cal D}_\mu \Psi_R+\bar \Psi_R \overleftarrow{\cal D}_\mu \Phi^\dagger {\cal D}_\mu \Psi_L\big{)}
\label{Wilson} \, ,
\end{eqnarray}
where $b^{-1}=\Lambda_{UV}$ is the UV-cutoff. We denote with  $\Psi_L=(u_L\,\,d_L)^T$ and $\Psi_R=(u_R\,\,d_R)^T$ the fermion  iso-doublets.
The Yukawa and Wilson-like terms are given by Eqs.~(\ref{Yukawa}) and (\ref{Wilson}), respectively. 
The latter is a six-dimensional operator multiplied by $b^2$ for dimensional reasons. 
The Yukawa coupling and the Wilson-like parameter are denoted by $\eta$ and $\rho$, respectively.
The scalar field  $\Phi=(\phi,-i\tau^2 \phi^*)$ is a $2\times2$ matrix with $\phi$  an iso-doublet 
of complex scalar fields. It obeys a quartic scalar potential denoted by the term  ${\cal V}(\Phi)$ of eq.~(\ref{LPHI})
where  $\mu_0^2$ and $\lambda_0$  are, respectively, the (bare) values for the squared mass and the self-interaction 
coupling constant of the scalar field. 
Moreover $F_{\mu\nu}^{a}$ is the field strength for the gluon field ($A_\mu^a$ with $a=1,2,\dots, N_c^2-1$). 
Finally, the covariant derivatives are given by: 
\beq
{\cal D}_\mu=\partial_\mu -ig_s \lambda^a A_\mu^a \, , \qquad
\overleftarrow{\cal D}_\mu =\overleftarrow{\partial}_\mu +ig_s \lambda^a A_\mu^a \, ,\label{COVG}
\eeq
A study of the unification of electroweak and strong interactions based on the above proposal has been presented in 
Ref.~\cite{Frezzotti:2016bes}. 
On-going work on the toy-model has been reported in Ref.~\cite{Capitani:2016ycp}.

\section{Symmetries and properties of the model}

The toy-model respects Lorentz, gauge,  and $C$, $P$, $T$ and $CPF_2$ symmetries (see Ref.~\cite{Frezzotti:2014wja}). 
Moreover it enjoys an exact symmetry under the global transformations $\chi_L$ and $\chi_R$ defined as:

\begin{equation}
\begin{array}{lcl}
\chi_L : \tilde{\chi}_L \otimes (\Phi \rightarrow \Omega_L \Phi), & &\chi_R : \tilde{\chi}_R 
\otimes (\Phi \rightarrow \Omega_R \Phi), \\
\hspace*{-1.0cm} \text{with~~~ } \tilde{\chi}_L: \Psi_L \rightarrow \Omega_L \Psi_L,   & & 
\tilde{\chi}_R: \Psi_R \rightarrow \Omega_R \Psi_R,  \\
\hspace*{0.6cm} \bar{\Psi}_L \rightarrow  \bar{\Psi}_L \Omega_{L}^{\dagger}, & & 
\hspace*{0.65cm} \bar{\Psi}_R \rightarrow  \bar{\Psi}_R \Omega_{R}^{\dagger},\\
\hspace*{-1.1cm} \text{where~~ } \hspace*{0.6cm} \Omega_L \in SU(2)_L, & &\hspace*{0.65cm} \Omega_R \in SU(2)_R. \\
 & &\\
 & &\\
\end{array}\vspace*{-0.9cm}
\label{eq:chi-sym}
\end{equation}
The toy-model~(\ref{SULL}), similarly to the LQCD case, is power-counting renormalizable with counter-terms constrained 
by the exact symmetries of the Lagrangian. In particular, thanks to the {\it exact} $\chi \equiv \chi_L \otimes \chi_R$ symmetry, owing to the 
inclusion of the scalar field in the Wilson term, there is no power divergent fermion mass terms, unlike to the Wilson-LQCD case. 
However the pure fermionic chiral transformations, $\tilde\chi \equiv \tilde\chi_L \otimes \tilde\chi_R$, 
do not constitute a symmetry of ${\cal L}_{\rm{toy}}$  due to the presence of the Yukawa and Wilson terms (for non-zero values of  
$\eta$ and $\rho$).

The physical implications of the toy-model depend crucially on the phase, Wigner or Nambu-Goldstone (NG), 
of the scalar potential ${\cal V}(\Phi)$. Following the line of argument of Ref.~\cite{Frezzotti:2014wja} it can be shown that 
$\tilde\chi$-symmetry enhancement takes place in the Wigner phase at a critical value of the Yukawa coupling. 
In fact by working in a way analogous to Ref.~\cite{Bochicchio:1985xa} one can get the renormalised 
Schwinger-Dyson equation (SDE) under $\tilde\chi_{L}$ transformations\footnote{Thanks to parity symmetry a similar equation holds for
the $\tilde\chi_{R}$ transformations.}:

\begin{equation} 
 \partial_\mu \langle Z_{\partial\tilde{J}}\tilde J^{L,i}_\mu(x) \,\hat {\cal O}(0)\rangle\! = 
 \!\langle \tilde\Delta_{L}^i \hat {\cal O}(0)\rangle\delta(x) -
({\eta- \overline\eta}) \,\langle {O_{Yuk}^{L,i}}(x)\,\hat {\cal O}(0)\rangle +{\ldots}+{\mbox{O}(b^2)}\, ,
\label{ren_SDE}
\end{equation}
in which the operator mixing under renormalisation of the $d$=6 operators with the two $d$=4 ones has been taken into account and  the current 
(the four-divergence of which is renormalised by $Z_{\partial\tilde{J}}\equiv Z_{\partial\tilde{J}}(\eta;g^2_s,\rho,\lambda_0)$) is 
defined by:
 \begin{equation}
\tilde J_\mu^{L\,, i}= \bar \Psi_L\gamma_\mu\frac{\tau^i}{2}\Psi_L -\frac{b^2}{2}\rho\Big{(}\bar \Psi_L\frac{\tau^i}{2}\Phi 
{\cal D}_\mu \Psi_R - \bar \Psi_R\overleftarrow {\cal D}_\mu\Phi^\dagger\frac{\tau^i}{2} \Psi_L\Big{)}\, .
\label{eq_current}
\end{equation}
Notice that thanks to the $\chi$-symmetry discretisation effects in Eq.~(\ref{ren_SDE}) are of $\mbox{O}(b^2)$ while the 
ellipses stand for possible contributions owing to possible NP operator mixing.   
The SDE of  Eq.~(\ref{ren_SDE}) becomes a WTI at a critical value of the Yukawa coupling, $\eta=\eta_{cr}(g^2_s,\rho,\lambda_0)$,
obtained by  $\eta_{cr}(g^2_s,\rho,\lambda_0)-\bar\eta(\eta_{cr}; g^2_s,\rho,\lambda_0)=0$. In this case 
$\tilde\chi$-symmetry restoration occurs, up to discretisation effects of $\mbox{O}(b^2)$, scalars get decoupled from quark 
and gluons, fermion mass is expected to vanish,  and  Eq.~(\ref{ren_SDE}) becomes:
\begin{equation}
 \partial_\mu \langle Z_{\partial\tilde{J}}\tilde J^{L,i}_\mu(x) \,\hat {\cal O}(0)\rangle\! = \!\langle \tilde\Delta_{L}^i \hat {\cal O}(0)\rangle\delta(x) 
 +{\mbox{O}(b^2)}\label{SYMCH} \, ,
\end{equation}
In the Wigner phase no spontaneous symmetry breaking (SSB) effect takes place, so the operator mixing is expected to follow 
perturbation theory arguments; as a consequence there are no ellipses in Eq.~(\ref{SYMCH}). In the NG phase instead,  a  
$\tilde\chi$SSB effect is expected to occur triggered by residual cutoff effects of $\mbox{O}(b^2)$, yielding new 
operator mixing terms of NP nature. In that case it is {\it conjectured} that Eq.~(\ref{ren_SDE}) takes the form:
    
\begin{equation}
  \partial_\mu \langle Z_{\partial\tilde J}\tilde J^{L,i}_\mu(x) \,\hat {\cal O}(0)\rangle_{\eta_{cr}} = 
\langle \tilde\Delta_{L}^i \hat {\cal O}(0)\rangle_{\eta_{cr}}\delta(x)+C_1\Lambda_s\langle 
{[ \overline \Psi_L \frac{\tau^i}{2}{\cal U} \Psi_R+\mbox{h.c.}]} \hat {\cal O}(0)\rangle +{\mbox O}(b^2)
\label{eq_wti_ng}
\end{equation}
where ${\cal U}$ is a dimensionless non-analytic function of $\Phi$ given by
\beq
{\cal U} = \frac{\Phi}{\sqrt{\Phi^\dagger \Phi}}=
\frac{v+\sigma+i\,\overrightarrow{\tau}\,\overrightarrow{\pi}}
{\sqrt{(v+\sigma)^2+\overrightarrow{\pi}\,\overrightarrow{\pi}}}\, .\label{U}
\eeq
The RGI term $ C_1\Lambda_s \bar{\Psi}_L \frac{\tau^i}{2}{\cal U} \Psi_{R}$ is $\chi_L \otimes \chi_R$ 
invariant\footnote{Note that a mass term 
of the form $[\bar{\Psi}_L \Psi_R + \bar{\Psi}_R \Psi_L]$ is not invariant under $\chi_L \otimes \chi_R$ transformations.}  
and is well defined only in the NG phase in which $\langle\Phi\rangle=v\neq 0$. $\Lambda_s$ stands for the 
scale of strong interactions that in our simulation setup (see next section) is identified with $\Lambda_{QCD}$.  
\

\section{Lattice simulations and results}
In this preliminary numerical study of the toy-model we have performed lattice simulations 
in the quenched approximation, where gauge and scalar fields can be generated independently.  
The verification or falsification process of the NP mechanism for fermionic mass generation 
is totally unaffected by the present choice to carry out simulations within  the (computationally cheap) quenched fermion approximation.   
We have employed naive Dirac fermions for which the $\chi_L \otimes \chi_R$ symmetry is exact.  
We have used the symmetric covariant derivative, $\tilde{\nabla}_{\mu}$,  throughout 
because with this choice the Wilson-like action term has symmetry properties
(see~\cite{BSMLAT17_1}, sect.~2) such that, even in the presence of fermion doublers, the value of $\eta_{cr}$ is unique.
In order to avoid exceptional configurations due to the possible presence of fermionic zero modes the twisted mass 
term, $i\mu_Q \bar \Psi \gamma_5\tau^3 \Psi$,  has been added in the lattice action (see Ref.~\cite{Frezzotti:2000nk}). 
The soft $\chi_L \otimes \chi_R$ symmetry breaking owing to the presence of the twisted mass term is eliminated in the limit 
$\mu_Q \rightarrow 0$.  
For full discussion of the lattice setup we refer the reader to the companion contribution at this conference~\cite{BSMLAT17_1}. 
\begin{figure}[t]
   \centering
   \subfigure[An example of the behaviour of the correlation function 
   $C_{\tilde{J}\tilde{D}}(x) \equiv \langle \tilde{J}_{0}^{V,\, 3} (x) \tilde{D}^{S,\, 3}(0)\rangle$ against the Euclidean time 
   for several values of $\eta$ at a certain value of $b\mu_Q=0.0224$.]%
             {\includegraphics[width=0.485\textwidth]{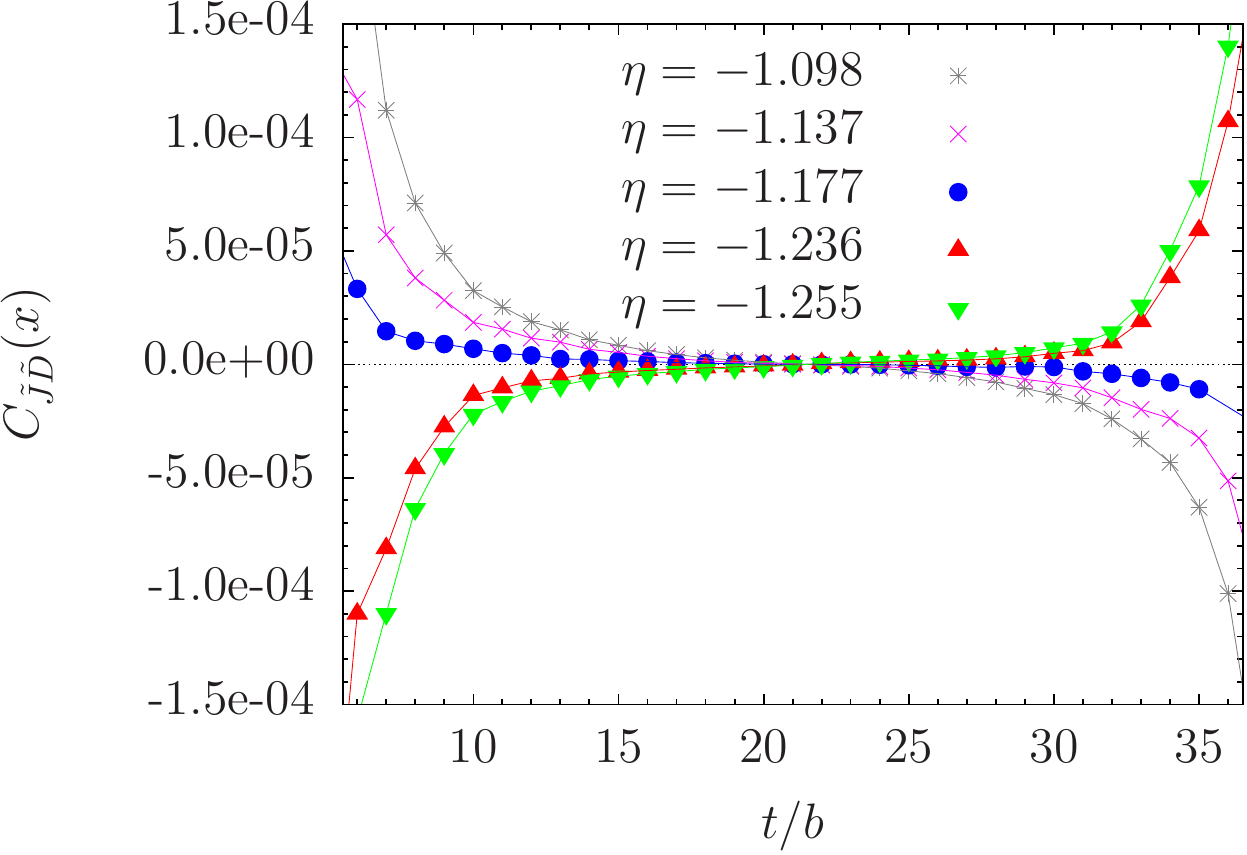}} \hfill
   \subfigure[Extrapolation of the ratio of correlation functions defined in Eq.~(\ref{r_WI}) 
   with respect to $\eta$. Results shown here have already been determined in the limit $\mu_Q \rightarrow 0$. Red-square symbol 
   indicates our estimate for $\eta_{cr}$.] %
             {\includegraphics[width=0.460\textwidth]{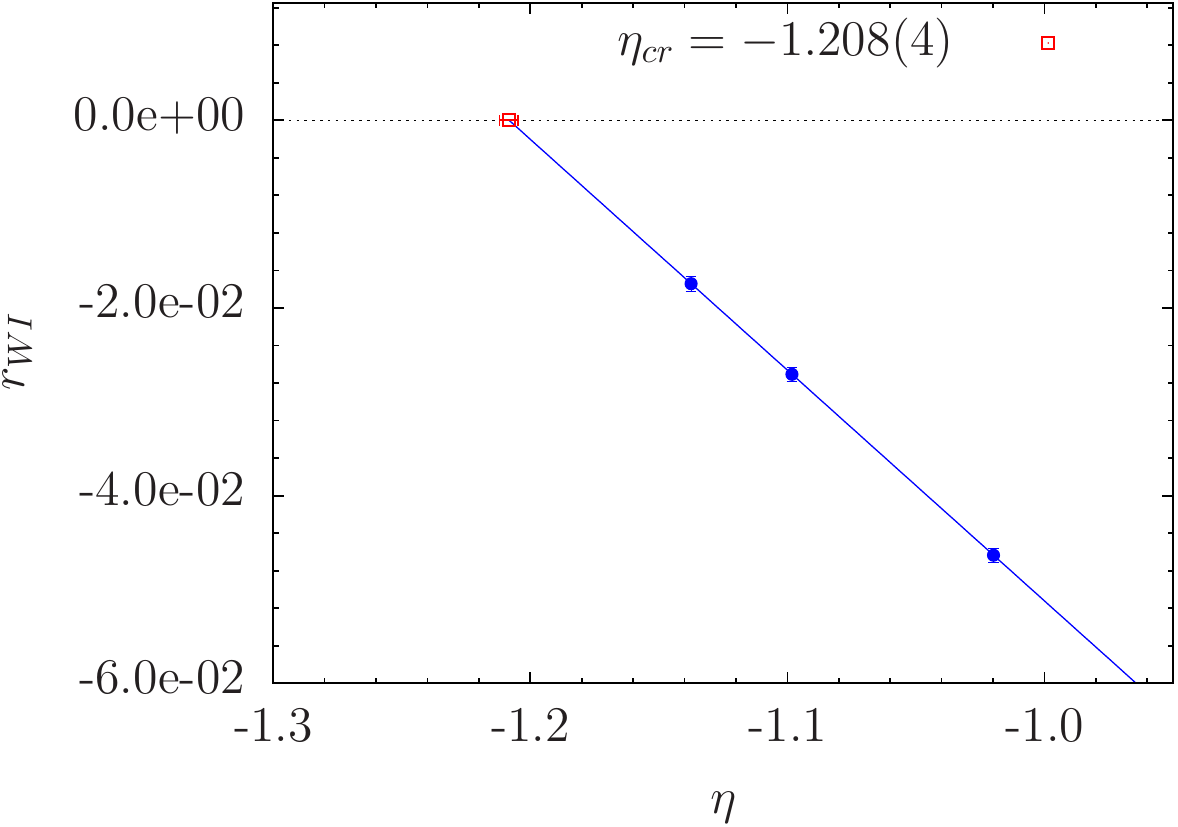}}
\caption{Results concerning the determination of the critical Yukawa coupling in the Wigner phase.}
\label{Wigner}             
\end{figure}

In these proceedings we present a preliminary status of the simulations and analysis of the results. 
We have performed simulations on a lattice volume $16^3 \times 40$
at one value of the gauge coupling ($\beta=5.85$) which corresponds to a lattice spacing of about $a = 0.123$ fm. 
Our lattice scale is given by $r_0 = 0.5$ fm determined in quenched LQCD in Refs~\cite{Guagnelli:1998ud} and \cite{Necco:2001xg}.  
For simulations in the Wigner and NG phases we keep fixed the value of the Wilson parameter ($\rho=1.961$),  the renormalised 
values of the $\sigma$-mass and the renormalised scalar coupling, i.e. 
$r_0^2m^2_{\sigma} = 1.276(6)$ and $\lambda_R = \frac{m_{\sigma}^2}{2v_{R}^2} = 0.4377(31)$. 
The statistics are 240 gauge $\times$ scalar  
configurations for several values of the Yukawa coupling, 
$\eta$, and at least three values of the twisted mass parameter, $\mu_Q$ for 
each value of $\eta$. 
For noise reduction we have used locally smeared scalar fields
in the lattice action.

\subsection{Determination of the critical Yukawa coupling in the Wigner phase}
In order to avoid  unnecessary contributions in the SDEs due to the presence of the twisted mass regulator in our lattice 
action, we employ the vector combination of $L$-handed and $R$-handed 
isotriplet currents, which obeys the following renormalized SDE (for $x\neq 0)$: 
\begin{equation}
 \partial_\mu \langle Z_{\tilde{J}}\tilde J^{V,3}_\mu(x) \, \tilde{D}^{S,3}(0)\rangle\! = 
({\eta- \eta_{cr}) \,\langle \tilde{D}^{S,3}}(x)\, \tilde{D}^{S,3}(0) \rangle +{\mbox{O}(b^2)} \label{SDE_V3}
\end{equation}
where we have defined:
\begin{eqnarray}
\hspace*{-0.cm}\tilde{J}_{0}^{V,\, 3} (x) &=& \tilde{J}_0^{L, \,3}(x) + \tilde{J}_0^{R, \,3}(x),  \nonumber \\
\hspace*{-0.cm} \tilde{D}^{S, \,3}(x) &=& \bar{\Psi}_L(x) \left[\Phi, \frac{\tau^{3}}{2}\right] \Psi_{R}(x) -  
\bar{\Psi}_R(x) \left[\frac{\tau^{3}}{2}, \Phi^{\dagger}\right] \Psi_{L}(x) 
\end{eqnarray}
and 
\beq
\hspace*{-0.cm} \tilde{J}_0^{L/R, \,3}(x) = \frac{1}{2} \left[\bar{\Psi}_{L/R}(x-\hat{0}) 
\gamma_0 \frac{\tau_3}{2} U_{0}(x-\hat{0}) \Psi_{L/R}(x) + 
\bar{\Psi}_{L/R}(x) \gamma_0 \frac{\tau_3}{2} U_{0}^{\dagger}(x-\hat{0}) \Psi_{L/R}(x-\hat{0}) \right].
\eeq

In the Wigner phase at $\eta=\eta_{cr}$ the correlation function 
$C_{\tilde J \tilde D}(x_0)  \equiv \sum_{\vec{x}} \langle 
\tilde J_0^{V,3}(x) \tilde D^{S,3}(0) \rangle $
is expected to vanish thanks to the restoration of the $\tilde\chi$--symmetry.
This behaviour can be noticed, as a tendency, by looking at the data in Fig.~\ref{Wigner}(a),
where the correlator $C_{\tilde J \tilde D}(x_0)$ is shown for several values
of $\eta$ at a certain value of $b\mu_Q = 0.0224$ (in lattice units). The vanishing
of $\lim_{\mu_Q \to 0} C_{\tilde J \tilde D}(x_0)$ at $\eta=\eta_{cr}$ implies, in
the absence of massless particles (which we explicitly check in our simulations), that all 
the on-shell matrix elements of $\tilde J_0^{V,3}$ must vanish in the same limit.

These remarks in turn suggest to determine $\eta_{cr}$ by looking at the renormalized
SDE of vector-$\tau^3$ $\tilde\chi$ transformations, namely
\beq
\partial_\lambda \tilde J_\lambda^{V,3}(x) = k_{\tilde J} (\eta-\eta_{cr}) 
\tilde D^{S,3}(x)  + {\rm O}(b^2) \; ,  \quad k_{\tilde J} = 
Z_{\partial \tilde J}^{-1} \frac{\eta-\bar\eta}{\eta-\eta_{cr}}
\eeq
with $k_{\tilde J}$ analytic in $\eta$ at $\eta=\eta_{cr}$ and O(1) (see~[7] about
$Z_{\partial \tilde J}$). 
This being an operator equation (with the form of a Ward Identity at $\eta=\eta_{cr}$) that holds 
on-shell for arbitrary intermediate states, it looks convenient to study the ratio 
\beq
r_{WI}(x_0) = 
\frac{\partial_0 \sum_{\vec{x}}\, \langle \tilde J_0^{V,3}(x) D^{S,3}(0) \rangle}{
      \sum_{\vec{x}}\, \langle  D^{S,3}(x)  D^{S,3}(0) \rangle} =
k_{\tilde J} (\eta-\eta_{cr}) + {\rm O}(b^2) \; . \label{r_WI}
\eeq
Indeed taking the average of $r_{WI}(x_0)$ over a $x_0$--window where only
few low-lying states contribute to the correlators in the ratio one gets a
quantity,  
\beq 
r_{WI}^{[\tau_1,\tau_2]}(\eta,\mu_Q) \equiv \frac{1}{\tau_2 - \tau_1} 
\sum_{x_0=\tau_1}^{\tau_2} r_{WI}(x_0;[\tau_1,\tau_2]) \; ,
\eeq
with reduced statistical noise and small O($b^2\Lambda_s^2$) deviations from 
$k_{\tilde J} (\eta-\eta_{cr})$. In particular, if $\eta_{cr}$ is determined by imposing the 
condition
\beq
r_{WI}^{[\tau_1,\tau_2]}(\eta = \eta_{cr};\mu_Q =0) = 0
\eeq
for an {\em appropriate time window $[\tau_1,\tau_2]$ kept fixed in physical units} 
at different lattice spacings, the O($b^2\Lambda_s^2$) cutoff effect in eq.~(\ref{r_WI}),
and the resulting one on the estimate of $\eta_{cr}$ at each $\beta$, by
construction will scale nicely towards zero as $b^2 \to 0$, thereby having no impact on
the properties of the {\em critical model that are established in the continuum limit}. 

The extrapolation of $r_{WI}^{[\tau_1,\tau_2]}(\eta,\mu_Q)$ to $\mu_Q=0$ is
easy in the Wigner phase, where absence of spontaneous symmetry breaking of
$\chi$-symmetry~\footnote{The study of the present toy model in the Wigner phase 
is possibly the first example in the literature of a {\em local} field theory where
confinement due to strong interactions takes place in the absence of spontaneous
chiral symmetry breaking.} and parity invariance entail an analytic dependence of 
$r_{WI}$ on $\mu_Q^2$, which happens to be numerically small and comparable to the 
statistical errors in the explored $\mu_Q$-range ($b\mu_Q = 0.0224, 0.0316, 0.0387$). 

The resulting values
of $r_{WI}^{[\tau_1,\tau_2]}(\eta;\mu_Q =0)$, for $[\tau_1,\tau_2] = [1.72,2.21]$~fm 
are shown in Fig.~\ref{Wigner}(b). Our preliminary result for the critical value of the Yukawa coupling determined in
this way at $\beta=5.85$ is $\eta_{cr}=-1.208(4)$.

\subsection{Dynamically generated fermion mass in the NG phase}
In the NG phase the $\chi_L \otimes \chi_R$ symmetry is broken to the $\chi_V$-symmetry. Moreover, at $\eta_{cr}$ the 
$\tilde\chi_L \otimes \tilde\chi_R$ symmetry, according to our conjecture, 
is expected to be spontaneously broken due to O$(b^2)$ effects. 
In Ref.~\cite{Frezzotti:2014wja} it has been argued that in the NG phase the local effective 
action density  of the model\footnote{The scalar potential here, $V_{\mu^2_\Phi<0}(\Phi)$, is written in terms of 
the renormalised parameters $\mu^2_\Phi$ and 
$\hat{\lambda}$. In the expression~(\ref{eq_Gamma_NG}) one could add one or more kinds of kinetic term of ${\cal U}$ that 
are proportional 
to $\Lambda_s$. However, for $v \gg \Lambda_s$ which is the typical regime for our mechanism these terms will 
be negligible.} 
reads:

\begin{equation}
\Gamma^{NG} 
= \frac{1}{4}(F\cdot F)+\bar Q \slash D Q + 
\frac{1}{2} \rm{Tr}\left[\partial_{\mu} \Phi^{\dagger}\partial_{\mu} \Phi\right] + V_{\mu^2_\Phi<0}(\Phi)
 + (\eta-\eta_{cr})(\bar \Psi_L\langle\Phi\rangle \Psi_R+{\mbox{h.c.}})+c_1\Lambda_s(\bar \Psi_L {\cal U} \Psi_R+{\mbox{h.c.}}). 
 \label{eq_Gamma_NG}
\end{equation}
We also note that in the NG phase the Wilson-like term gets effectively 
a form analogous to the one of the Wilson term in Lattice QCD. Indeed by 
setting $r = bv \rho$ (with $v$ the scalar field vev) and neglecting quantum
field fluctuations the Wilson-like term in the toy model lattice action
can be rewritten in the form
 $$ {\cal L}_{Wil}^{QCD}(\Psi, A) = -\frac{br}{2}  \left( \bar{\Psi}_L D^2 \Psi_R + \text{h.c.} \right). $$
Simulations in the NG phase are performed by employing the same values for the set of the parameters ($\beta$, $\lambda_R$, $\rho$)
and the lattice volume as  in the Wigner phase. 
\begin{figure}[tp]
   \centering
   \subfigure[Bare values of the fermion mass, $2 r_0 m^{WTI}$, in units of $r_0 =0.5$ fm at several values of $\eta$. 
   All results have been extrapolated to zero twisted mass. The straight line passing from the points is 
to guide the eye. We indicate the rough numerical estimate for the non-perturbatively generated fermion mass at $\eta_{cr}$ and 
the value of $\eta$, namely, $\eta^{*}$ at which the fermion mass vanishes.]%
             {\includegraphics[width=0.460\textwidth]{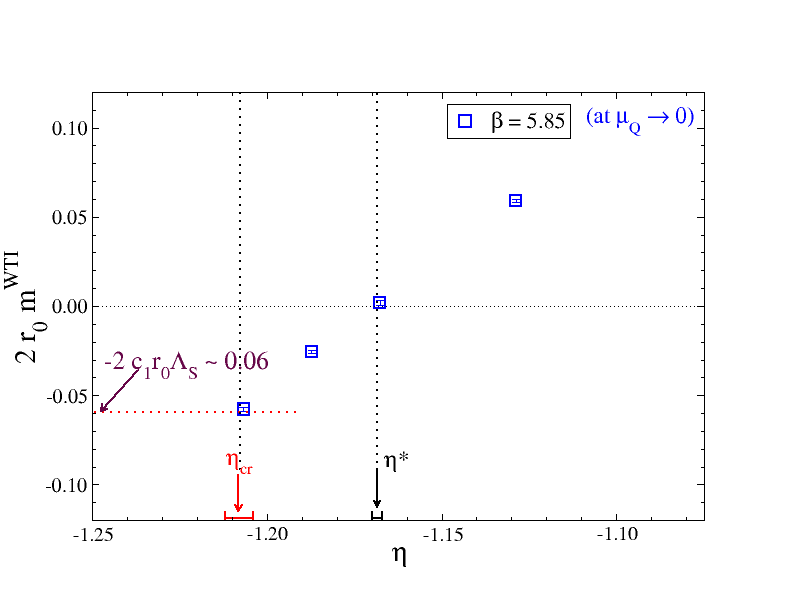}} \hfill 
   \subfigure[Results for $M_{\text{PS}}^2$ in units of $r_0^2$ at the same values of $\eta$ as in the left panel. 
   All results have been 
   extrapolated to zero twisted mass. We explicitly indicate in physical units the estimates for 
$M_{\text{PS}}$ at $\eta_{cr}$ and $\eta^{*}$ (see the text for details).]%
             {\includegraphics[width=0.46\textwidth]{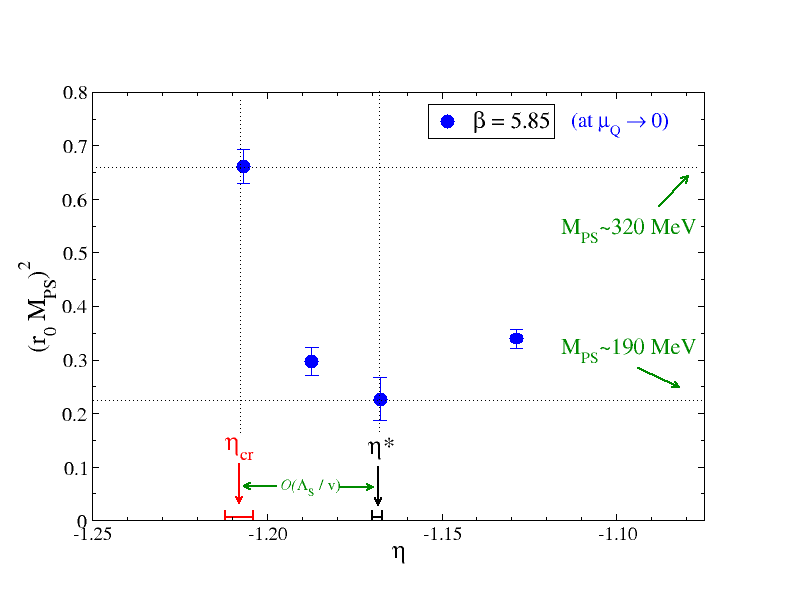}}
\caption{Results (preliminary) for $2 r_0 m^{WTI}$ and $(r_0 M_{\text{PS}})^2$ in the NG phase at several values of the Yukawa coupling.}
\label{NG}             
\end{figure}

The effective quark mass (in the $\mu_Q=0$ limit) can be read off from the
axial $\tilde\chi$ WTI, e.g.
\begin{equation}
2 m^{WTI}=\dfrac{b^{-1}\partial_0\langle 0|\widetilde J^{A \pm}_0|M_{PS^\pm}\rangle}{\langle 0|P^\pm|M_{PS^\pm}\rangle} 
\label{eq_mWTI}
\end{equation}
where 
$$\widetilde J^{A\pm}_0 (x) = \bar{\Psi}(x-\hat{0}) 
\gamma_0 \gamma_5\dfrac{\tau_1 \pm i \tau_2}{2} U_{0}(x-\hat{0}) \Psi(x) + 
\bar{\Psi}(x) \gamma_0 \gamma_5 \dfrac{\tau_1 \pm i \tau_2}{2} U_{0}^{\dagger}(x-\hat{0}) \Psi(x-\hat{0})$$  is the one-point-split current associated to the 
fermionic  ($\tilde\chi$) axial transformations and    
$P^{\pm}(x) = \bar{\Psi}(x) \gamma_5 \dfrac{\tau_1 \pm i \tau_2}{2} \Psi(x)$ is the pseudoscalar density. 

In Fig.~\ref{NG}(a) we show results for the bare quark mass (multiplied by a factor of two) 
in units of $r_0$ against the Yukawa coupling. 
The results  have been obtained  
using Eq.~(\ref{eq_mWTI}) at several values of $(\eta, \mu_Q)$. 
For each value of $\eta$ a linear extrapolation to $\mu_Q = 0$ has been performed.
Small deviations from linearity are possible and their impact is presently
under study by extra simulations at further $\mu_Q$ values and more elaborate
fits.
At $\eta=\eta_{cr}$, where the Yukawa quark mass term gets cancelled, 
the $m^{WTI}$ is expected to be equal to the conjectured quark mass of NP origin, $c_1\Lambda_s$. 
As it can be seen from that figure, based on our preliminary data, a rough estimate of the bare quark 
mass\footnote{The work for the quark mass renormalisation is on-going. 
The method is described in the companion contribution~\cite{BSMLAT17_1}.} 
in $r_0$ units is $-2 r_0 c_1 \Lambda_s \simeq 0.06$. Passing now to Fig.~\ref{NG}(b) where $(r_0 M_{\text{PS}})^2$ is shown against 
the Yukawa coupling we notice that at $\eta=\eta_{cr}$ the corresponding value for the pseudoscalar mass is rather large (of about 320 MeV or larger).  
We would also like to draw the attention to an interesting feature which occurs at the value of the Yukawa coupling, 
namely $\eta^{*}$, 
at which $m^{WTI}$ vanishes. With the help of the 
effective action density of Eq.~(\ref{eq_Gamma_NG}) one can deduce that, 
defining $m^{WTI} \equiv (\eta^{*} - \eta_{cr}) v + c_1 \Lambda_s = 0$ entails
$\eta^{*} = \eta_{cr} - c_1 \Lambda_s / v$.
Our data gives evidence that $\eta^{*} - \eta_{cr} \neq 0$ which further supports the conclusion 
that the dynamically generated quark mass is 
non-zero\footnote{Subsequent work following our presentation at Lattice 2017 has provided
further numerical results at two lattice spacings, which strenghtens 
the evidence in favour of the dynamical fermion mass generation mechanism
that is discussed here, see~\cite{bsm_forthcoming}.}.        

\section{Summary and further developments}
We have discussed a toy-model that exemplifies a novel NP 
mechanism proposed in Ref.~\cite{Frezzotti:2014wja} for dynamical fermion mass generation. The fundamental property  
of the mechanism consists in the enhancement of the QCD symmetries in such a way that fermion masses emerge 
in a {\it natural} way~\cite{tHooft:1979rat}, being independent from the Yukawa interaction and the scalar field.   
Thanks to the NP character of the mechanism the physical implications and predictions of the associated toy-model 
can be tested with the help of simulations on the lattice. 
We have presented  preliminary results based on simulations in the 
quenched approximation at one value of the lattice spacing. Our results for  
the dynamically generated effective fermion mass and the associated pseudoscalar meson mass in the NG phase, 
barring cutoff effects, are of O$(\Lambda_s)$. Since the presentation at the conference we have  performed 
more simulations at the present lattice spacing  and  improved our methods of analysis. 
We have also carried out simulations at a second value of the lattice spacing in order to be able to check the 
scaling behaviour both of the fermion mass and the 
pseudoscalar meson mass. All these results that show rather smooth scaling properties will be presented soon in~\cite{bsm_forthcoming}.    
\vspace*{0.7cm}

\noindent{\bf Acknowledgements}\\
We acknowledge support from INFN, via the convention INFN-Cineca 
which made available to us the CPUtime for carrying out numerical
simulations on Galileo, Marconi A1 and Marconi A2 clusters. 
Support from the Sino-German CRC110 research network is also gratefully
acknowledged.

\clearpage
\bibliography{lattice2017}

\end{document}